\pdfoutput=1
\documentclass[11pt]{article}

\usepackage{acl2023}

\usepackage{times}
\usepackage{latexsym}

\usepackage{latexsym}
\usepackage{graphicx}
\usepackage{subcaption}
\usepackage{graphicx} % For \resizebox
\usepackage{booktabs} % For cleaner tables
\usepackage[T1]{fontenc}

\usepackage[utf8]{inputenc}

\usepackage{microtype}

\usepackage{inconsolata}
\usepackage{booktabs} 
\usepackage{algorithm}
\usepackage{algorithmicx}
\usepackage{algpseudocode}

\usepackage{enumitem}
\usepackage{verbatim}
\usepackage{multirow}
\usepackage{diagbox}

\usepackage[normalem]{ulem}
\newcommand{\zhou}[1]{{\color{black}{#1}}} % zhou edit
\newcommand\blfootnote[1]{%
  \begingroup
\renewcommand\thefootnote{}\footnote{#1}%
  \addtocounter{footnote}{-1}%
  \endgroup
}

\title{A Framework for Cost-Effective and Self-Adaptive LLM Shaking and Recovery Mechanism}

\author{\normalsize
Zhiyu Chen$^{1}$, Yu Li$^{2}$, Suochao Zhang$^1$, Jingbo Zhou$^{2,*}$, Jiwen Zhou$^{1,*}$, Chenfu Bao$^1$, Dianhai Yu$^1$\\
$^1$Baidu Inc., $^2$Baidu Research\\
\{chenzhiyu, liyu31, zhangsuochao, zhoujingbo, zhoujiwen, baochenfu, yudianhai\}@baidu.com
}

\begin{document}
\maketitle

\blfootnote{$^*$Corresponding authors.}

\begin{abstract}
\zhou{As Large Language Models (LLMs) gain great success in real-world applications, an increasing number of users are seeking to develop and deploy their customized LLMs through cloud services. Nonetheless, in some specific domains, there are still concerns regarding cost and trade-offs between privacy issues and accuracy. In this study, we introduce a cost-effective and self-adaptive LLM shaking tuning and recovery mechanism, named CypherTalk. With carefully designed horizontal and vertical shaking operators, we can achieve comparable accuracy results with SOTA privacy-preserving LLM schemes using Cryptography-based or Differential Privacy-based methods. Experiments also show that with the CypherTalk framework, users can achieve reliable accuracy when using optimized shaking operator settings. To our best knowledge, this is the first work that considers cost, and trade-off between model utility and privacy in LLM scenarios. }
\end{abstract}

\section{Introduction}

 Large Language Models (LLMs) have made significant progress in natural language understanding and generation. As LLMs and Models as a Service (MaaS) continue to evolve and gain popularity, more and more users are turning towards utilizing MaaS offerings from cloud service providers to implement their customized LLMs.
Nevertheless, in the practical deployment of LLMs, privacy and security concerns remain significant challenges. Fine-tuning LLMs for downstream tasks on cloud servers is a crucial method for enhancing their applicability to specific domains. However, the data employed in this fine-tuning process often possesses a heightened sensitivity, particularly in sectors such as healthcare, finance, and government. Therefore, how to enhance the efficiency of fine-tuning for LLM while simultaneously protecting user privacy in the cloud platform context emerges as a challenging research area currently.

\zhou{
Recent research efforts have been directed towards devising privacy-protected fine-tuning solutions for Large Language Models (LLMs), broadly categorizing into two distinct approaches. The first category encompasses crypto-based methods, such as Homomorphic Encryption (HE) \cite{chen2022x}, and Secure Multi-party Computation (MPC) \cite{li2022mpcformer,dong2023puma}. These techniques, while effective in maintaining privacy, often suffer from excessive computational overhead, making them less practical for fine-tuning and inference on private data. In contrast, the second category comprises methods based on Differential Privacy (DP), as exemplified in studies like \cite{li2021large, du2023dp}. These DP-based approaches inherently prioritize privacy. However, they frequently incur a trade-off, typically manifesting as considerable reductions in accuracy, especially under standard DP configurations. This trade-off highlights a fundamental challenge in achieving an optimal balance between privacy preservation and model accuracy within the study of privacy-protected LLM fine-tuning.}

%There have some some research attempts to address the privacy-protected LLM fine-tuning solution. It can be generally divided into two categories.  The first category is the crypto-based methods like Homomorphic Encryption(HE) and Secure Multi-party Computation(MPC). However, thye usually have too much computation overload to perform fine-tuning and inference on private data.  Compared with DP-based methods, However, algorithms based on differential privacy(DP) inherently prioritize privacy, often resulting in significant reductions in accuracy under standard DP configurations. 

%\begin{figure}[!t]
%    \centering    
%    \begin{center}       %\includegraphics[width=0.50\textwidth]{pics/cipherllm1.png}
%        \caption{Overview of CypherTalk}
%        \label{fig:overview} 
%    \end{center}
%\end{figure}
 % On one hand, based on differential privacy theory, appropriately disrupting the intrinsic data distribution within the embedding can achieve the goal of protecting the privacy of training data.
% perturbed embedding can act as a form of regularization, forcing the model to adapt to noisy inputs and suppress over-fitting, thereby benefiting the enhancement of the model's generalization ability.

% On the other hand, according to differential privacy theory, appropriately disrupting the intrinsic data distribution within the embedding can achieve the goal of protecting the privacy of training data. By jointly optimizing these two objectives, we can obtain an fine-tuned LLM that is both privacy-preserving and with high quality of evaluation.

\zhou{Our insight is that the shaking-based privacy-preserving mechanism provides an alternative way to fine-tune the LLM. This technique involves integrating an optimized noise module within the LLM's representation layer during fine-tuning. Such an approach can achieve both enhanced privacy protection in the cloud service setting, and preserving model quality. Therefore, we introduce \textbf{CypherTalk}, a cost-effective and self-adaptive shaking and recovery mechanism. Within the CypherTalk framework, fine-tuning processes can securely outsource computational tasks to cloud platforms without compromising on predictive accuracy. This method effectively addresses the dual objectives of enhancing data privacy and ensuring high model performance in cloud-based computations.}
We summarize the key contributions of our work as follows:

\begin{itemize}
    \item We propose a cost-effective framework that allows developers to train customized privacy-preserving LLMs on a cloud platform, providing reliable services to end users.
    \item We design a shaking and recovery mechanism, constructing an end-to-end, parameter-configurable privacy-protected framework for LLM fine-tuning and inference.  
    % \item We conduct several evaluations based on multiple instruction fine-tuning datasets and tasks to assess the impact of the framework on the model's generalization performance. The goal is to minimize performance loss under strict differential privacy.
    \item Our approach demonstrates the effectiveness and efficiency of our method over its competitors. We also conduct security analysis under the proposed mechanism. 
\end{itemize}

\section{Related Work}

To protect customer data privacy and the proprietary of LLM weights, existing works can be generally divided into two categories, which are cryptographic methods and differential privacy method.

Cryptographic methods usually incorporate homomorphic encryption (HE)\cite{gentry2009fully} and secure multiparty computation (MPC) \cite{shamir1979share,yao1986generate} in the training and inference processes of Large Language Models (LLMs). Notably, nGraph-HE \cite{boemer2019ngraph} provides a HE backend to the deep learning models. THE-X \cite{chen2022x} has recently introduced a practical method enabling pre-trained transformer models to perform inferences under HE by incorporating an approximation of LayerNorm into the fine-tuned model. Recent investigations have explored the projection of LLMs within the MPC framework, as evidenced in studies like \cite{li2022mpcformer,dong2023puma}. These approaches typically require a substantial computational cost.

% In contrast to HE, another branch of cryptographic methods is MPC which can provide a safe solution by keeping data and model weights private during inference but with an unacceptable high inference cost\cite{evans2018pragmatic}. To accelerate the process, MPCFormer \cite{li2022mpcformer} offers a private inference service utilizing transformer models which is structured as a two-party computation and expedites the inference process by approximating functions like GeLU and Softmax. Another recent MPC framework work is PUMA \cite{dong2023puma}, which leverages replicated secret sharing and optimizes computational costs of the transformer model by approximating polynomials on CPUs. While these cryptographic methods offer robust security guarantees, they still entail considerable computational and communication overheads. 

%the following up related work wil lbe revised
% \zhoucom{The related work should describe the general framework, and tell the difference between the important paper and our methods. \{Done\}  }
Differential Privacy methods mainly include Local Differential Privacy(LDP) and DP-SGD. DP-forward provides perturbation methods in the forward pass to speed up DP-SGD latency and reach a higher accuracy at the same time. \cite{yu2021differentially} propose a meta-framework for differential private fine-tuning methods. The experiments show that differentially private adaptations perform well in utility, privacy, and the computational and memory cost of private training. In \cite{mai2023split}, the authors provide an innovative framework that splits the model to execute the token embedding layer on the client side with very little computational cost. The authors in \cite{majmudar2022differentially} proposed a simple, lightweight perturbation mechanism that applies differential privacy at the decoding stage.

% Another privacy regularization\cite{mireshghallah2021privacy} method jointly utilizes differential privacy and discriminator to optimize the privacy-utility trade-off with acceptable computation overhead. With the huge requirement of trade-off between utility, privacy and cost, how to jointly optimize the triangle is a recent challenge among researchers.

However, both of these two categories have their own disadvantages. For example, cryptography-based methods cannot support private tuning because of the complexity of adopting HE or MPC. Even though DP-based methods can support private tuning, the accuracy of DP-based methods depends largely on the privacy parameters, and as the layer becomes larger, the accuracy drops dramatically. 

\section{CypherTalk Framework}
\label{sec:CypherTalk}

\zhou{There are mainly four stages in the CypherTalk framework: 1) Key Generation, 2) Key Implantation, 3) Private tuning process,  and 4) Private inference Process. In the rest of this section, we explain our proposed pipeline in more detail.}

\subsection{Phase I: Key Generation Process}\label{sec:keygen}

\zhou{In the key generation phase, the generated plays a pivotal role in modulating (i.e. shaking) the model, governing the encryption and decryption of the data in and out of the model.  This shaking process encompasses two distinct orientations: horizontal and vertical shaking. Depending on the specific requirements, this shaking process can be executed once or multiple times. To facilitate this, a pair of unique keys is employed, denoted as $key: {vs_{key}, hs_{key}}$, which correspond to the vertical shake and horizontal shake, respectively. The Algorithm \ref{alg:key_gen} delineates the procedure for generating these keys, ensuring separate and distinct key creation for both horizontal and vertical directions.}

\begin{algorithm}
\caption{Key Generation }
\begin{algorithmic}
\label{alg:key_gen}
\State\textbf{Input:} $vs_{keys}=\{vs_n, vs_{ops}[], vs_{opts}[]\}$
\State$hs_{keys} = \{hs_{tab}\}$
\State\textbf{Output:} $keys = \{vs: vs_{keys}, hs: hs_{keys}\}$ 
\State\textbf{Process:}
\Statex Step1: Vertical key generation:
    \For{$i \gets 0$, $vs_n$}
        \State $o \gets rand(0, len(vs_{ops})$
        \State $vs_{op} \gets vs_{ops}[o]$
        \State $vs_{opt} \gets vs_{op}.keygen() + vs_{opts}[o]$
    \EndFor
\Statex Step2: Horizontal key generation:
\Statex $seq \gets range(0, len(vocab))$
\Statex $hs_{tab}\gets permute(seq)$ 
\Statex Step3: Key Combination:
\Statex $key=\{vs_{key}, hs_{keys}\}$
\end{algorithmic}
\end{algorithm}

\textbf{Vertical Key Generation}.
 \zhou{The vertical key in our framework is denoted as $vs_{key}: \{vs_n, vs_{op}, vs_{opt}\}$. During the generation of vertical keys, $vs_{n}$ specifies the number of vertical transformations to be applied. For each transformation, the algorithm randomly selects a vertical operation function, $vs_{op}$, and subsequently generates a corresponding key based on the chosen vertical operator. The term $vs_{op}$ represents the specific transformation function employed, with the experimentally identified functions being ${addv, inflate, tilt, dx-fixp, gaussian, laplace}$. For comprehensive details of these functions, refer to Table \ref{tab:vs_ops}. The component $vs_{opt}$ pertains to the parameters of the transformation. It primarily involves the random matrix used in the transformation function and the associated hyperparameter settings.}

\begin{table}[t]
\centering
\caption{Vertical Shaking Operators}
\label{tab:vs_ops}
\scalebox{0.8}{
    \begin{tabular}{c|l|l}
    \hline
    \textbf{VS. ops} & \textbf{Key generation} & \textbf{Key implantation} \\
    \hline
    Addv & $Vec(rand(n)\times \Delta)$ & $E + Addv$ \\
    Inflate & $diag(\mathcal{N}(0, \sigma^2)\times \Delta)$ & $E * Inflate$ \\
    Tilt & $T(Vec(\mathcal{N}(0, \sigma^2))\times \Delta)$ & $E * Tilt$ \\
    $D\chi-fixp$ & $Vec(\gamma(k, \Theta))\times \Delta)$ & $E+D\chi-fixp$\\
    Gaussian & $Vec(\mathcal{N}(\epsilon, l))\times \Delta)$ & $E+Gaussian$\\
    Laplace & $Vec(lap(\epsilon, l))\times \Delta)$ & $E+Laplace$\\
    \hline
    \end{tabular}
}
\end{table}

\textbf{Horizontal Key Generation}.  \zhou{The horizontally generated key functions as a mapping table, correlating old and new IDs. This concept, termed ``horizontal shaking'', essentially involves the random permutation of an ordered sequence. The structure of a horizontal key is defined as $hs_{key}: {hs_{tab}}$, where $hs_{tab}$ represents a mapping table constructed through a random process. This approach underpins the fundamental methodology of horizontal shaking in our framework.

}

\subsection{Phase II: Key Implantation Process}
\begin{figure*}[!ht]
    \centering    
    \begin{center}
        \includegraphics[width=1\textwidth]{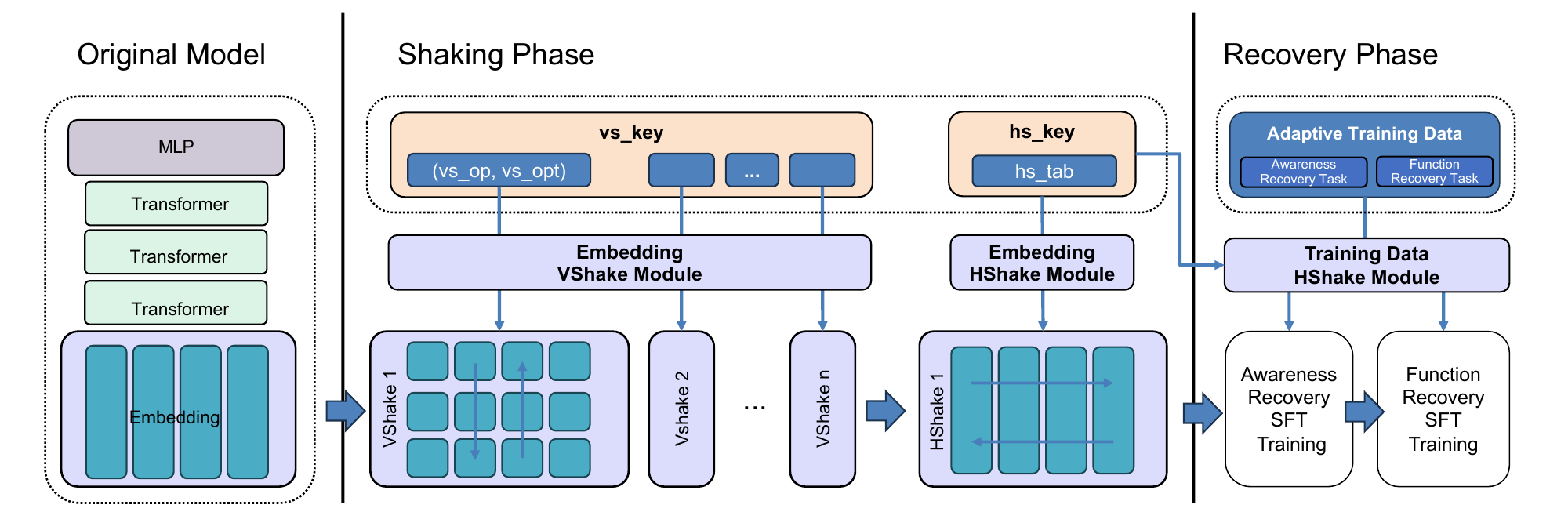}
        \caption{Key Implantation Process}
        \label{fig:implant} 
    \end{center}
\end{figure*}

\zhou{Key Implantation involves embedding a user-generated key into the language model, enabling the model to be encrypted. At the same time, the client can explicitly modify the model to accept encrypted input and output encrypted answers as part of an operational process. This includes two sub-stages: the Model Shaking Stage and the Model Adaptation Stage, both of which are integral to the overall key implantation and embedding process.  The Model Shaking Stage involves the deliberate distortion of the model's embedding layer, in accordance with the specifications of the Vertical Key and Horizontal Key as discussed in Section \ref{sec:keygen}. Following this, the Model Adaptation Stage is employed to reinstate the original effectiveness and capabilities of the model, which might have been compromised due to the shaking. Given that the shaking process substantially alters the model's embedding layer, a consequent decline in performance is often observed. Thus, the Model Adaptation Stage is crucial for implementing training strategies aimed at countering the functional impacts of shaking, a process also named as model recovery. Through the synergistic operation of these two stages, the model is enabled to adjust to the alterations in its representation induced by shaking, while simultaneously striving to maintain a performance level that closely aligns with that of the unmodified original model.}

\begin{algorithm}
    \caption{Key Implantation Process}
    \label{alg:key_implant}
    \begin{algorithmic}
        \State\textbf{Input:} Original Model $M=\{O, T, E\}$, 
        \State {$vs_{keys}=\{vs_n, vs_{ops}[], vs_{opts}[]\}$ $hs_{keys} = \{hs_{tab}\}$ $AdaptData=\{X, Y\}$} 
        \State\textbf{Output:} shaked model $M'$
        \vspace{1mm}
        \For{$i \gets 0$,$vs_n$}
            \State\textbf{Step1: Model Repression Layer Shaking}
            \State Vertical shake embedding layer of Model, Use algorithm \ref{vertical_shaking}
            \State $vs_{op}=vs_{ops}[i]$ \& $vs_{opt}=vs_{opts}[i]$
            \State $M'=vs\_shake(M', vs_{op}, vs_{opt})$
            \State Horizontal shake embedding layer of Model, Use algorithm \ref{horizontal_shaking}
            \State $M'=hs\_shake(M', hs_{tab})$ %\COMMENT{} 
            \State\textbf{Step2: Adaptive Recover Training}
            \State Awareness Recovery Training, Use Algorithm \ref{alg:private_tune}
            \State $privacy\_train(M', AdaptData(X, X), hs_{keys})$
            \State Function Recovery Training, Use Algorithm \ref{alg:private_tune}
            \State $privacy\_train(M', AdaptData(X, Y), hs_{keys})$
        \EndFor
\end{algorithmic}
\end{algorithm}

%During the use of large language models in the cloud, customized model will be fine-tuned through interfaces on the cloud. During the process, plaintext interaction method poses a privacy risk to users when using large models on the cloud. Specific privacy risks include cloud transmission risks, operator theft risks, and snooping risks between tenants. Our CypherTalk mechanism performs above operation in a permuted way so that the privacy and security can be guaranteed between clients and server. 

\subsubsection{Model Shaking Phase}
\zhou{The Model Shaking Phase involves shaking the parameters of the representation layer of the mode, enabling the transformation and encryption of the model's input and output as dictated by the user-specified key. The representation layer, in this context, pertains to the model's input and output layer. For large language models, the input layer specifically refers to the weights of the word vectors in the word embedding layer. Conversely,  the output layer is associated with the decoding weights of the last layer in the model. For a task-specific model, such as a classification language model, because the hidden layer of the output is not directly used in the text generation process, it is only connected to the classifier of the upper layer to achieve its task-specific classification work. Shaking can be directed only at the input layer without transforming the output layer.  The shaking operation involves the manipulation of the weights in the representation layer, with two primary objectives: 1) To alter the representation of the original model, and 2) To differentiate the weights of the transformed model from those of the original. To fulfill these objectives, two model shaking are employed: \textit{Model Horizontal Shaking} and \textit{Model Vertical Shaking}. }

%The Model Shaking Phase, refers to the shaking of the parameters of the representation layer of the model, so that the expression of the input and output of the model can be transformed in the way defined by the key specified by the user, so as to quickly realize the controlled change of the expression mode of the model. The representation layer of the model refers to the input layer of the model, and for the large language model, it refers to the word vector weight of the word embedding layer of the model. The output layer, for large language models, refers to the decoding weight of the MLP layer at the top layer of the model. For a task-specific model, such as a classification language model, because the hidden layer of the output is not directly used in the text generation process, it is only connected to the classifier of the upper layer to achieve its task-specific classification work. Shaking can be directed only at the input layer without transforming the output layer. The shaking operation refers to the transformation of the weights of the representation layer of the model, and the objectives are: 1. To transform the representation of the original model. 2. Make the weights of the transformed model different from those of the original model. In order to achieve the above two goals, there are two types of model shaking: \textit{Model Horizontal Shaking} and \textit{Model Vertical Shaking}.

\zhou{\textbf{Model Vertical Shaking} refers to the process of altering the weights within the model's representation layer, which is shown in Algorithm \ref{vertical_shaking}. The primary aim of this process is to disrupt the association between the transformed and original vectors, thereby obscuring the relation between the transformed vector and its original counterpart. Vertical shaking typically involves multiple iterations, each employing a transformation operator and a unique key. The key for vertical shaking is defined as $vs_{key}:{vs_{op}, vs_{opt}}$, where $vs_{op}$ denotes the transformation function utilized. Further details about the Vertical Key are elaborated in Section \ref{sec:keygen} and Algorithm \ref{alg:key_gen}. Upon determining the operator for a given round, the client implements the specified operation $vs_{op}$ on various word vectors within the model's representation layer. This process distorts and transforms the word vectors, ensuring that the transformed vectors are markedly distinct from their original forms. Such transformation is instrumental in obfuscating the correlation between the word vectors and their original counterparts. Importantly, this key is retained on the client side and remains inaccessible to the server. Vertical shaking can be executed one or more times, depending on client requirements, with the specific number of transformations being dictated by $vs_{n}$.}

\begin{algorithm}
    \caption{Model Vertical Shaking}
    \label{vertical_shaking}
    \begin{algorithmic}
        \Statex\textbf{Input:} Language Model $M=\{O, T, E\}$ 
        \Statex{$vs_{keys}=\{vs_{op}, vs_{opt}\}$}
        \Statex\textbf{Output:} Vertical shaked model $M'$
    
        \Statex Get weight from word embedding layer
        \Statex $W_E \gets E$ \& $W_E'=vs_{op}(W_E, vs_{opt})$
        \Statex Get weight from output embedding layer
        \Statex $W_O \gets O$ \& $W_O'=vs_{op}(W_O, vs_{opt})$
        \Statex Update model with new weights
        \Statex $M'=\{ O(W_O'),T,E(W_E')\}$
    \end{algorithmic}
\end{algorithm}

\zhou{\textbf{Model Horizontal Shaking}, as delineated in Algorithm \ref{horizontal_shaking}, pertains to the transformation of weights within the model's embedding layer. This process occurs along the axis of the word list, or the horizontal dimension of word vectors, hence the term ``horizontal shaking''. During horizontal shaking, word vectors in the embedding layer are rearranged into column units, effectively altering their positions. The key for horizontal shaking is denoted as $hs_{key}:\{hs_{key}\}$, representing a mapping table between old and new IDs. This mapping table is generated by randomizing and swapping elements within an ordered sequence. In the context of data encryption, the IDs of the input data are transposed according to the mapping specified by $hs_{key}$, resulting in encrypted word IDs. Conversely, during decryption, the word IDs output by the model are decoded using the $hs_{key}$, thereby retrieving the original word IDs. This method ensures that the word vectors are effectively shuffled, contributing to the model's security.}

\begin{algorithm}
    \caption{Model Horizontal Shaking}
    \label{horizontal_shaking}
    \begin{algorithmic}
        \Statex\textbf{Input:} Language Model $M=\{O, T, E\}$ 
        \Statex{$hs_{keys} = \{hs_{tab}\}$} 
        \Statex\textbf{Output:} Horizontal shaked model $M'$
    \State Get weight from word embedding layer
    \State $W_E \gets E$ \& $W_E'=permute(W_E, hs_{tab})$
    \State Get weight from output embedding layer
    \State $W_O \gets O$ \& $W_O'=permute(W_O, hs_{tab})$
    \State Update model with new weights
    \State $M'=\{ O(W_O'),T,E(W_E')\}$
    \end{algorithmic}
\end{algorithm}

\subsubsection{Model Adaptation Phase}
\zhou{The model adaptation phase refers to the adaptive training operation to enable the original model to accommodate the altered representation layer, to reinstate the functions and efficacy of the original model. The adaptation phase of the model consists of two phases: \textit{Awareness Recovery Training} and \textit{Functional Recovery Training}.}

\zhou{\textbf{Awareness Recovery Training} is a stage aimed at enabling the model to understand the transformed representation layer. During this phase, the model is trained to recognize and adapt to the distribution of the newly transformed word vector space. To facilitate this understanding, simple training tasks, such as recapitulation training tasks, can be employed. Specifically, in scenarios where the recovery training dataset is based on Question Answering (QA), a dataset format of ${X=Q, Y=Q}$ can be utilized for the training during this recovery phase. Generally, models designed for specific tasks tend to adapt more readily compared to generative models. Similarly, models with fewer layers exhibit a higher ease of adaptation than those with numerous layers. }

\zhou{\textbf{Functional Recovery Training} is a stage dedicated to restoring the function of the model, enabling it to adjust to the alterations introduced by the shaking transformation. The primary objective is to ensure that the functionality or efficacy of the shaken model either matches or closely approximates that of the original model.  For instances where the recovery training involves a Question Answering (QA) dataset, a dataset format of ${X=Q, Y=A}$ is suitable for this phase. }

\subsection{Phase III: Private Tuning Process}
\label{sec:tuning}

\zhou{In the private tuning process, our mechanism tailors the shaking of private data according to the specific tuning tasks for each client. Before clients transmit their training samples to the server, sensitive attributes within the data are either removed or altered by a data processing module. Notably, this process involves encoding the data into tokenized IDs, which are then mapped onto a new space using  $hs_key$, as detailed in Algorithm \ref{alg:private_tune}. Subsequently, the server receives these encoded and shaken training samples and undertakes fine-tuning on the transformed training samples. This fine-tuning can optionally incorporate Parameter Efficient Fine-Tuning (PEFT) configurations, thereby enhancing the adaptability and efficacy of the process under different operational scenarios.}

% \zhoucom{this cloud-client setting should be explained in Section 3 Preliminaries.\{done\}}

% \subsubsection{Smoothness Assumption of CypherTalk} For shaking and recovery tuning, we assume that the target downstream task should have a similar distribution with implant auxiliary classifiers. For any two points $x, x' \in \mathcal{X}$, \zhoucom{what is the definition of $x, x'$, , and what does $f(x)$ mean? ans: $x$ is tokenizer} we have following inequality where
% L is a non-negative real number known as the Lipschitz constant, which provides an upper bound  rate of the function: 
% \begin{equation}
%     \|f(x)-f(x')\| \leq L\|x-x'\|
% \end{equation}
% \zhoucom{what does this assump mean? ans: from transfer learning

% \zhoucom{This paragraph is not clear, we have explain clearly how to conduct the Key-tuple Implant Mechanism. What is perturbation module? What is "random one-way perturbation table? "
% what does mean "we provide different tuning methods for perturbed input?"
% Which is loss is used? We cannot leave this choice to users...

% }

% \zhoucom{what do we use in the experiment? ans: CE loss}

% As shown in algorithm \ref{alg:tuning}, 
% \zhoucom{what does the mean of "reconstruct the permutation ids?" ans: recover}
\begin{algorithm}
\caption{Private Fune-tuning Mechamism}
\begin{algorithmic}
\label{alg:private_tune}
\Statex\textbf{Input:} $hs_{key}$, encoded\_model $M$, llm\_vocab $v$, input\_text $\{\mathcal{X},y\}$,  finetune\_opts $vs_{key}$  \vspace{1mm} 
% \item {$hs_{keys} = {hs_{tab}}$} \vspace{1mm} \newline

\Statex\textbf{Output:} $fine\_tuned\_model$ $M\prime$  \vspace{1mm} 

\Statex\textbf{Private tuning process:} \vspace{1mm} 
\Statex\textbf{Loop} \vspace{1mm}
\Statex\textbf{Step1: Privacy-aware Data Encryption} \vspace{1mm} 
\Statex \hspace{3mm}$X\_ids$ = encode$(v, input\_\mathcal{X})$ 
\Statex \hspace{3mm}$y\_ids$ = encode$(v, input\_y)$
\Statex \hspace{3mm}$P\_input = f(X\_ids,  hs_{key})$
\Statex \hspace{3mm}$P\_label = f(y\_ids,  hs_{key})$

\Statex\textbf{Step2: Private Fine-tuning Mechanism}

\Statex \hspace{3mm}\textbf{For} $freeze\_param\ in\  model.parameters():$ 

\Statex \hspace{8mm}$freeze\_param.requires\_grad = False$ 
\Statex \hspace{3mm}$l' = infer(M', \{P\_input, P\_label\}, opts)$ 
\Statex \hspace{3mm}$loss = loss\ (l', P\_label)$ 
\Statex \hspace{3mm}$delta = \Delta loss$ 
\Statex \hspace{3mm}$encoded\_model.w = model.w + \alpha \times delta$ 

\Statex\textbf{end Loop}
\end{algorithmic}
\end{algorithm}

\subsection{Phase IV: Private Inference Process}
\label{sec:inference}

\zhou{Leveraging the privately held key tuple settings, clients are enabled to process queries and obtain final results based on the server's response. In the computation process, the client utilizes the updated horizontal key, $hs_{key}$, to shake their tokenized input before dispatching it to the outsourced computation center. This privacy-centric model adeptly calculates the shaken embeddings and returns the resultant shaken ID set to the client. In the post-inference phase, the client applies her  $hs_{key}$ to decode the resulting label IDs. Throughout this entire procedure, both the input data and the results remain encrypted, ensuring the safety and integrity of the data under various circumstances. }
% \begin{algorithm}
% \caption{Private Inference Algorithm}
% \begin{algorithmicx}
% \label{alg:inference}
% \Statex \textbf{Input:} $queries$, $hs\_key$, llm\_vocab $v$,
% \Statex \hspace{5mm} shaked\_tuned\_model\ $M'$
% \Statex \textbf{Output:} $res$ \vspace{1mm}

% \Statex \textbf{Step1: Private calculating process:} \vspace{1mm}
% \Statex \textbf{At Server Side:}

% \Statex \hspace{3mm} Load the shaked\ tuned\ model\ $M'$.
% \Statex \hspace{3mm} Split customized classifier MLP layer $M_{c}$ to client and keep $M_{s}$ parts.

% \Statex \textbf{At Client Side:}
% \Statex \hspace{3mm} Client generates input embedding $ids_{input}$ using $v$ and $hs\_key$.
% \Statex \hspace{3mm} Client calculates private input embeddings $E_{hs}$ and send intermediate  $res_{inter}$ to server. \vspace{1mm}

% \Statex \textbf{Step2: Post inference process:} \vspace{1mm}
% \Statex \textbf{At Server Side:} 
% \Statex \hspace{3mm} Server receives intermediate value $res_{inter}$ and calculates the output results $res_{ids\prime}$.

% \Statex \hspace{3mm} Send the cipher result $res_{ids\prime}$. to clients.

% \Statex \textbf{At Client Side:}
% \Statex \hspace{3mm} Client receives the cipher result and decode using the $hs_{tab}$ to get the final results $y$. 
% \Statex
% \end{algorithmicx}
% \end{algorithm}

\section{Experiments}

\zhou{In this section, we present an experimental evaluation to demonstrate the performance of CypherTalk in terms of accuracy and cost, in comparison with state-of-the-art counterparts. Additionally, a security analysis is conducted to assess CypherTalk with its baseline models.}

%We investigate the following four research questions:

%\noindent• \textbf{RQ1.} How does our proposed CypherTalk perform when compared with the state-of-the-art?

%\noindent• \textbf{RQ2.} How does our proposed CypherTalk cost when compared with the state-of-the-art?

%\noindent• \textbf{RQ3.} How does CypherTalk perform when varying injection key type settings?

%\noindent• \textbf{RQ4.} How can different operators be used to provide the model safety?

% • RQ4. How do different components  XXX contribute to the overall performance?

% \zhoucom{here please give a paragraph to explain the main purpose of the experiments, what do we want to prove or verify？ like this :

% We investigate the following four research questions:

% • RQ1. How does our proposed ... perform when compared with the state-of-the-art XXX?

% • RQ2. How does XXX perform when varying XXX
% settings (e.g., embedding size)?

% • RQ3. How can XXX be used to provide the model safety?

% • RQ4. How do different components  XXX contribute to the
% overall performance?}

\subsection{Settings} 

\textbf{Datasets.} In order to compare with other state-of-the-art methods, we evaluate CypherTalk on the GLUE dataset. There are six tasks in this dataset which include Stanford Sentiment Treebank v2 (SST2), Quora Question Pairs(QQP), Corpus Corpus of Linguistic Acceptability (CoLA), Recognizing Textual Entailment (RTE), Stanford Question Answering Dataset(QNLI), Multi-Genre Natural Language Inference (MNLI)\cite{wang2018glue}.
% \zhoucom{should give the citation of the dataset}

\noindent \textbf{Baselines.} To validate the performance, we compared against the following state-of-the-art methods which are mainly divided into two categories. We choose PUMA and MPCFormer as crypto-based baselines, and  DP-forward, DP-SGD as DP-based method baselines. \newline
•MPCFormer\cite{li2022mpcformer}: MPCFormer utilizes Secure Multi-Party Computation(MPC) and Knowledge Distillation(KD) for privacy-preserving Transformer inference.\newline
• PUMA\cite{dong2023puma}: PUMA enables secure Transformer model inference by utilizing secret share scheme and designing approximations for expensive functions such as GeLU and softmax.\newline
• DP-SGD\cite{li2021large}:  DP-SGD enables privately training Transformers with a modest run-time overhead by clipping gradients.\newline
• DP-forward\cite{du2023dp}: DP-Forward perturbs embedding matrices in the forward pass of LMs to satisfy local DP requirements.

\noindent\textbf{Experimental Configurations.} We implemented our mechanisms and baselines in Python and performed our experiments on a cluster with NVIDIA Tesla V100 GPUs. We adopt ACC and Matthews Correlation Coefficient (MCC) on the GLUE dataset. The parameters of the DP-based method are chosen from the default upper bound in the corresponding papers. Table \ref{tab:sota_list} compares all methods including crypto-based and DP-based after performing three epochs tuning.

%However, these methods have different strengths and weaknesses across different dimensions, and we compare these differences on GLUE tasks. 
% \zhoucom{please explain three subsubsection in this paragraph, which are 

% \bf{Datasets}XXXX

% \bf{baselines} XXX

% \bf{Experimental Configurations}XXX

% }

% \begin{table*}[!h]
% \centering
% \caption{Parameters and functionality of State-of-the-Art Methods}
% \label{tab:sota_list}
% \begin{tabular}{|c|c|c|c|c|c|c|}
% \hline
% \multicolumn{2}{|c|}{Method} &Framework & Model  & Layers & Hidden & Fine-tuning \\ \hline
% \multirow{HE} & MPCFormer\cite{li2022mpcformer} &PyTorch\/ & BERT & 12 & 768 & No \\ \cline{2-7} 
%  & THE-X\cite{chen2022x} &PyTorch\slash SEAL& BERT & 12 & 1024 & No \\ \cline{2-7} 
%  & PUMA\cite{dong2023puma} &PyTorch\slash SPU & BERT & 12 & 1024 & No \\ \hline
% \multirow{DP} & DP-forward\cite{du2023dp}&PyTorch & BERT & 24 & 1024 & Yes \\ \cline{2-7} 
%  & DP-SGD\cite{li2021large}&PyTorch\slash Opacus & BERT & 24 & 1536 & Yes \\ \hline
% \multirow{Ours} & Adaptive Shaking \& Recover &Paddle & BERT & 24 & 1024 & Yes \\ \cline{1-7} 
% \end{tabular}
% \end{table*}

\begin{table*}[!h]
\centering
\caption{Parameters and functionality of State-of-the-Art Methods}
\label{tab:sota_list}
\begin{tabular}{|c|c|c|c|c|c|c|}
\hline
\multicolumn{2}{|c|}{Method} & Model  & Layers & Hidden & Fine-tuning & Inference \\ \hline
\multirow{3}{*}{HE} & MPCFormer\cite{li2022mpcformer} & BERT & 12 & 768 & No & Yes  \\ \cline{2-7} 
 & THE-X\cite{chen2022x} & BERT & 12 & 1024 & No & Yes \\ \cline{2-7} 
 & PUMA\cite{dong2023puma} & BERT & 12 & 1024  &  No & Yes  \\ \hline
\multirow{2}{*}{DP} & DP-forward\cite{du2023dp}& BERT & 24 & 1024 & Yes & Yes \\ \cline{2-7} 
 & DP-SGD\cite{li2021large} & BERT & 24 & 1536 & Yes & Yes \\ \hline
\multirow{1}{*}{Ours} & Adaptive Shaking \& Recover & BERT & 24 & 1024 & Yes & Yes \\ \cline{1-7} 
\end{tabular}
\end{table*}

% \zhoucom{I do not understand what this paragraph mean?}

% Since the previous works are all performed on BERT-series models, in order to make a fair comparison with previous SOTA privacy-preserving methods, we also implemented our mechanisms on Bert series model as our backbone as shown in table \ref{tab:sota_list}. We use the same dataset that presented in\cite{{li2022mpcformer}}\cite{chen2022x}\cite{dong2023puma}

% \textbf{Baselines.} 
% \zhoucom{we have to give brief explaination about what the baselines is, and its reference.}

% \textbf{Implementation.} We run our experiments on a cluster with NVIDIA Tesla V100 GPUs. We implemented our mechanisms and baseslines in Python. 

% \zhoucom{The first section following the "Setting Section" should be the subsection about "main experiment results"}

% \zhoucom{The reviewers do not care the  "Crypto-based vs DP-based utility" and  "Crypto-based vs DP-based Cost ". They only care "Our method vs other baselines"}

\subsection{CypherTalk vs Baseline Utility}
% \zhoucom{In the first paragraph of each section ,give a summary introduction of this section.}

\begin{table*}[htb]
\centering
\caption{Performance on GLUE benchmark. MCC and Match ACC are reported for CoLA and MNLI respectively. ACC is reported for the rest tasks. If not specified, experiments are performed with 3 epochs.}
\label{tab:test_accuracy}
\begin{tabular}{@{}cccccccc@{}}
\toprule
\diagbox{Method $ /\ $  Task} & \textbf{SST-2}  & \textbf{QQP} & \textbf{CoLA} & \textbf{RTE} & \textbf{QNLI} & \textbf{MNLI}\\ 
\midrule
Vanilla Bert model           & \textbf{0.93}          & \textbf{0.91}          & 0.60      & \textbf{0.71}          & \textbf{0.92}          & \textbf{0.85}\\
% THE-X  (Bert-Tiny)    &  0.82      &  0.83         & ----       &    0.58     &   0.78        &    0.68                  \\
PUMA (Bert-Base)   &  ----     &  ----      & \textbf{0.61}         & \textbf{0.70}          &  \textbf{0.91}       & ----     \\
MPCFormer (Bert-Base(Quad+2Quad))    &  \textbf{0.92}     &   \textbf{0.90}        & 0.50         & 0.55        &    \textbf{0.91}      &  \textbf{0.85}    \\
DP-SGD (Bert-Base, $\epsilon=8$)    &  0.83       &  0.66       &  \textit{0.54}         & 0.47         & 0.64           & 0.33     \\
DP-forward (Bert-Base, $\epsilon=8$)    &  0.87           &  0.86       &    0.50       & 0.53         & 0.88      &  0.79              \\
Ours (Bert-Base) @3epoch   & 0.92          & 0.88         & 0.57      & 0.63        & 0.91        & \textbf{0.85}               \\
%Ours (Bert-Base) @10epoch   & 0.92          & 0.88         & \textbf{0.59}      & 0.62        & \textbf{0.92}         & \textbf{0.85}               \\
Ours (Bert-Base) with adapt @10epoch & \textbf{0.93}     & \textbf{0.91}          & \textbf{0.59}          &  \textbf{0.69}           &  \textbf{0.91}       &   0.84        \\
% Ours (Bloom-560M)      &             &           &   0.81     &    0.64      &          &                \\ 
\bottomrule
\end{tabular}
\end{table*}

% Differential privacy based-methods provide a defense against re-identification attacks by adding randomized noise to the data. However, cumulative noisy may lead to increased error and decreased utility in the results. On the other hand, an adversary with access might be able to infer sensitive details about individuals, even in the presence of differential privacy mechanisms. We will discuss in more detail in section 5.5. \zhoucom{what does mean of this paragraph}

 %The key observations can be summarized as follows.
\zhou{We have performed several benchmarks to evaluate CypherTalk as shown in Table \ref{tab:test_accuracy}, the cryptography-based method can reach a similar accuracy level. From the table, we can also conclude that MPCFormer performs better than HE-based and DP-based methods when $\epsilon$ reaches 8\footnote{Among cryptography-based methods, we only reproduced MPCFormer due to compile environment limitations. Therefore, we cited the results from \cite{dong2023puma} on CoLA, RTE, and QNLI tasks.}. Our method is a little lower than the Vanilla Bert model but performs best among cryptography-based and DP-based methods. A possible reason behind this performance is that our method can find suitable noise to adapt to the corresponding dataset types. }

\zhou{Cryptography-based methods often underperform, primarily due to their limitation to only supporting the inference process, without the capability for fine-tuning. The constraints preventing these methods from facilitating the fine-tuning stage can be attributed to several factors: 1) The modification of traditional fine-tuning methods is too complex to accomplish; 2) Even with limited parties involved in the training process, the time complexity is hard to accept due to the huge volumes of computation and communication between involved parties.  }

\subsection{CypherTalk vs Baseline Cost}
\begin{figure*}[htb]
\centering

\begin{subfigure}[b]{0.47\textwidth}
    \includegraphics[width=\textwidth]{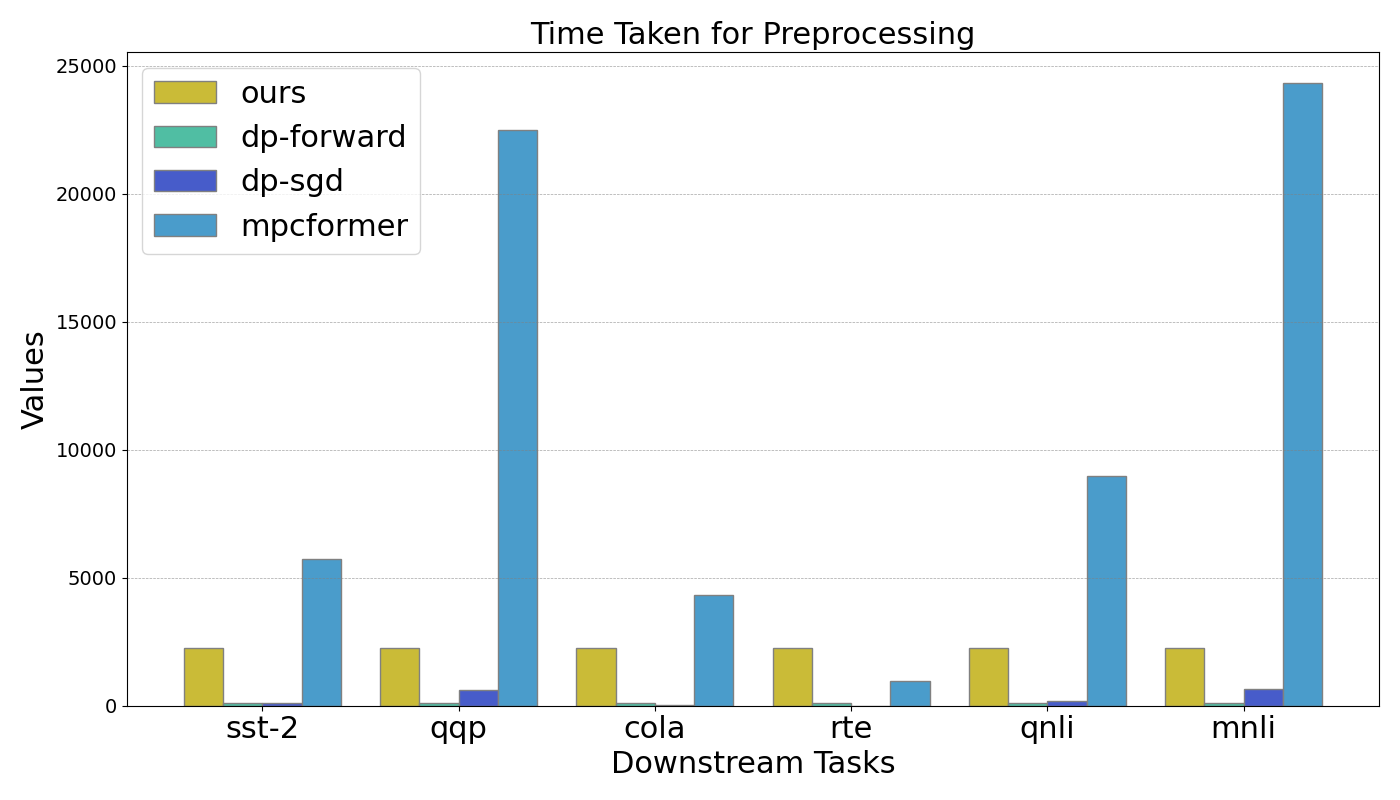}
    \caption{Time costs for pre-processing}
    \label{fig:pre-processing-time}
\end{subfigure}
\hfill % this will insert a non-breaking space that will push the subfigures to the edge of the text area
\begin{subfigure}[b]{0.47\textwidth}
    \includegraphics[width=\textwidth]{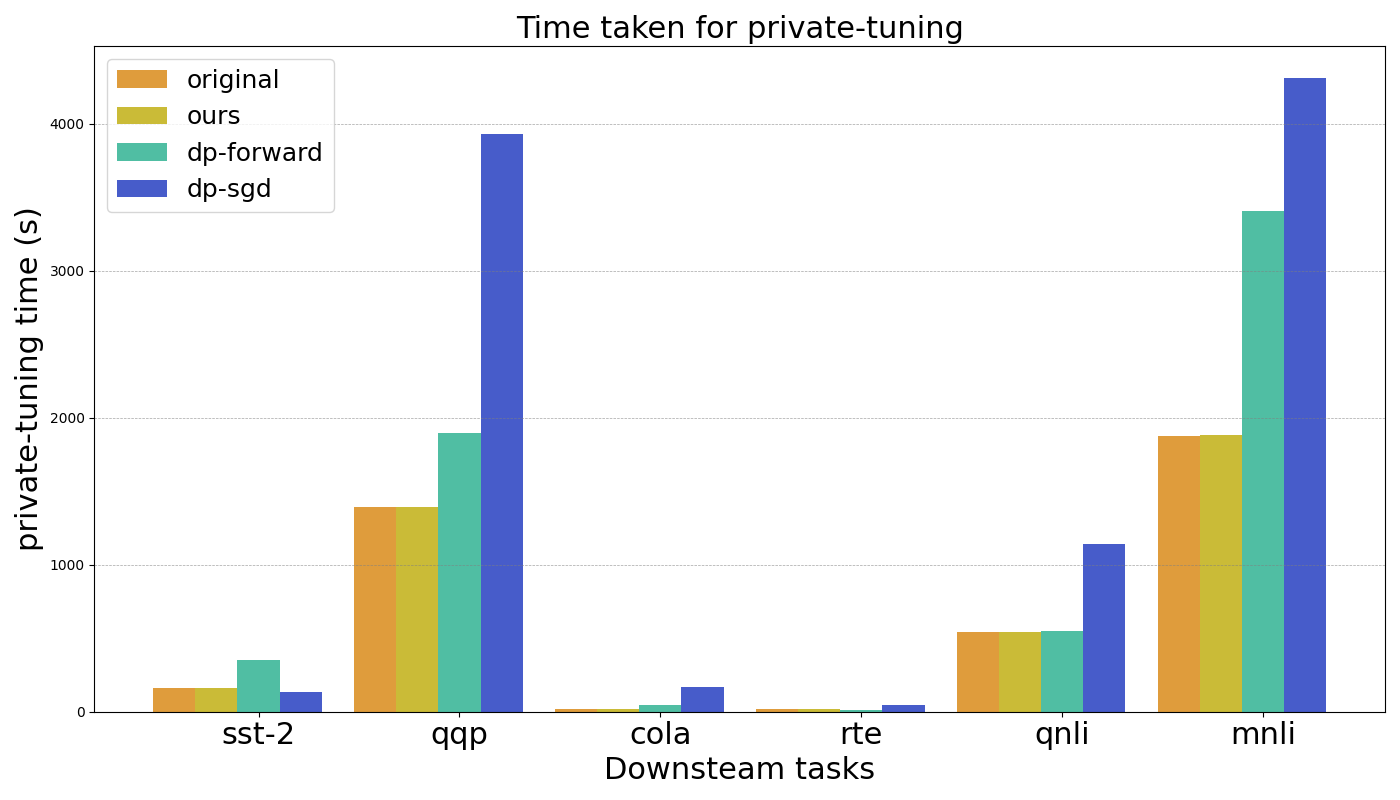}
    \caption{Time costs for tuning}
    \label{fig:tuning-time}
\end{subfigure}
\hfill
\caption{Time cost of different SOTA methods.}
\label{fig:cost}

\end{figure*}
As shown in Figure \ref{fig:cost}, the costs of pre-processing time are different between ours and other baselines. The pre-processing time cost of our method reminds same respect for different downstream tasks. However, MPCFormer uses knowledge distillation to help speed up inference time cost, but the pre-processing times cannot be ignored in order to achieve an acceptable evaluation accuracy of downstream tasks. The preprocessing of MPCFormer includes model fine-tuning and distillation. The pre-processing time cost of DP-SGD and DP-forward are mainly come from data preparation. 
Different with DP-based and MPC-based methods, the pre-processing time cost of CypherTalk is mainly come from the key-tuple implant which includes horizontal shaking and vertical shaking and recovery mechanism. %\zhoucom{The above the below discussion should be more about ChiperTalk, not about baselines.}

\zhou{Cryptography-based methods operate computation on ciphertext which ensures the privacy protection of each individual. As shown in Figure \ref{fig:pre-processing-time}, the time of pre-processing on the cryptography-based method is largest compared with DP-based and ours, because the fine-tuning and distillation are needed before the models are ready for use. From Figure \ref{fig:tuning-time} we can observe that the time cost for private tuning of DP-based is larger than CypherTalk because of the overheads to perform training and perturbed data or gradients for DP-based methods.}

\begin{figure}[htb]
\centering
\includegraphics[width=0.45\textwidth]{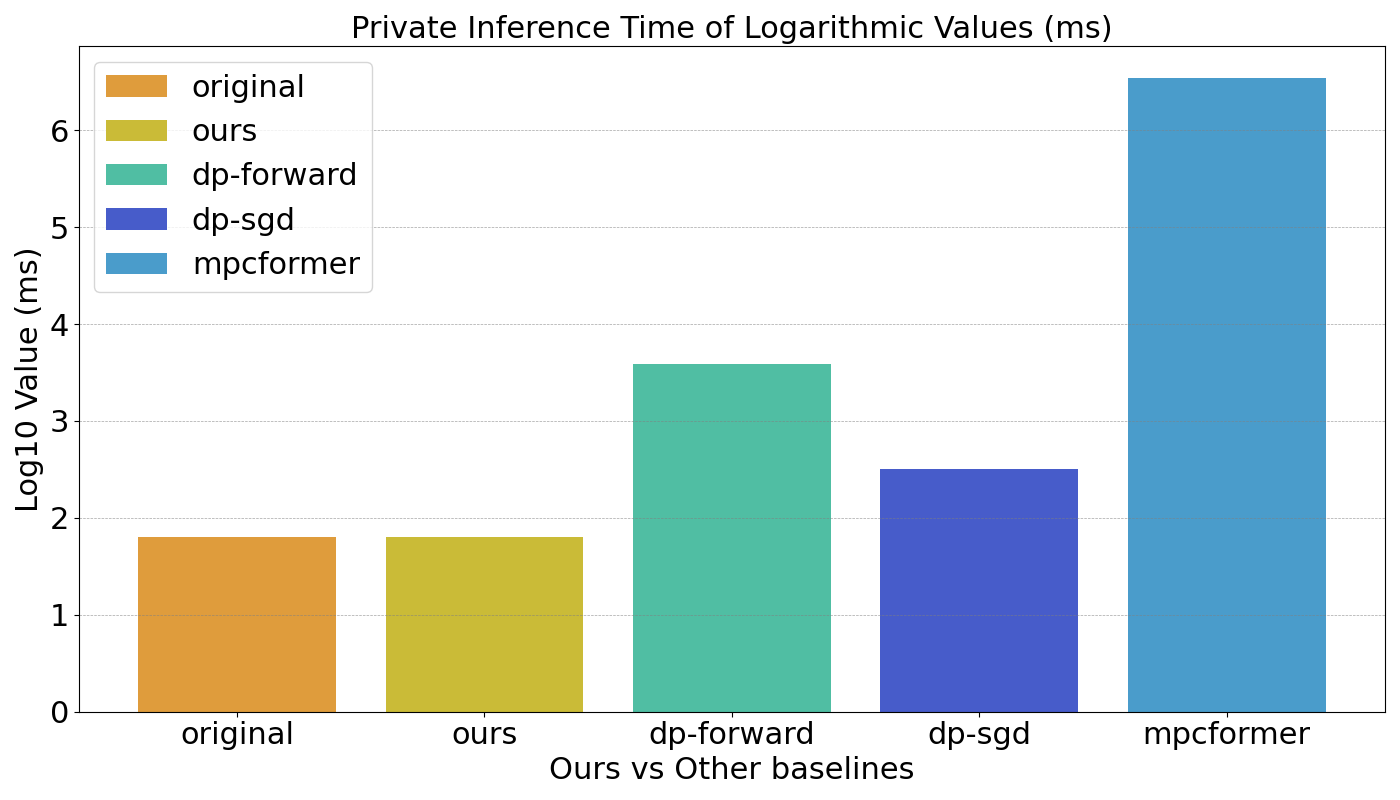}
\caption{Impact on inference time}
\label{fig:infer-impact}
\end{figure}

\zhou{From figure \ref{fig:infer-impact} we can observe that the inference time costs of cryptography-based methods is hundreds or even a thousand times than DP-based mechanisms and our CypherTalk. Because the inference time of MPC-based methods needs a lot of cryptography operation on plain text.}

%\begin{itemize}[leftmargin=*,itemsep=0pt,parsep=0pt]
%\item Cryptography-based methods operate computation on ciphertext which ensures the privacy protection of each individual, helps protect against insider threats or malicious entities within an organization. But the overhead of performing operations on encrypted data can be a significant drawback. As shown in \ref{fig:pre-processing-time}, the time of pre-processing on MPC-based method is largest compared with DP-based and ours, because the fine-tuning and distillation are needed before the models are ready for use. From figure \ref{fig:tuning-time} we can observe that the time cost for private tuning of DP-based is larger than ours because there are some overhead when performing training and perturbed data or gradients for DP-based methods.

%\item From figure \ref{fig:infer-impact} we can observe that the inference time costs of cryptography-based methods is hundreds or even a thousand times than DP-based mechanisms and our CypherTalk. Because the inference time of MPC-based methods needs a lot of cryptography operation on plain-text which is hard to accept in large-scale application. \zhoucom{what about our method?}
%\end{itemize}

\subsection{Impact of Different Noise Types}

In this section, we performed different noise operators as our injection key set on SST-2 downstream task. In this experiment setting, we use different noise operators to generate the $vs_{key}$ and output their shaken embeddings.   
\begin{figure}[htb]
\centering

\begin{subfigure}[b]{0.45\textwidth}
    \includegraphics[width=\textwidth]{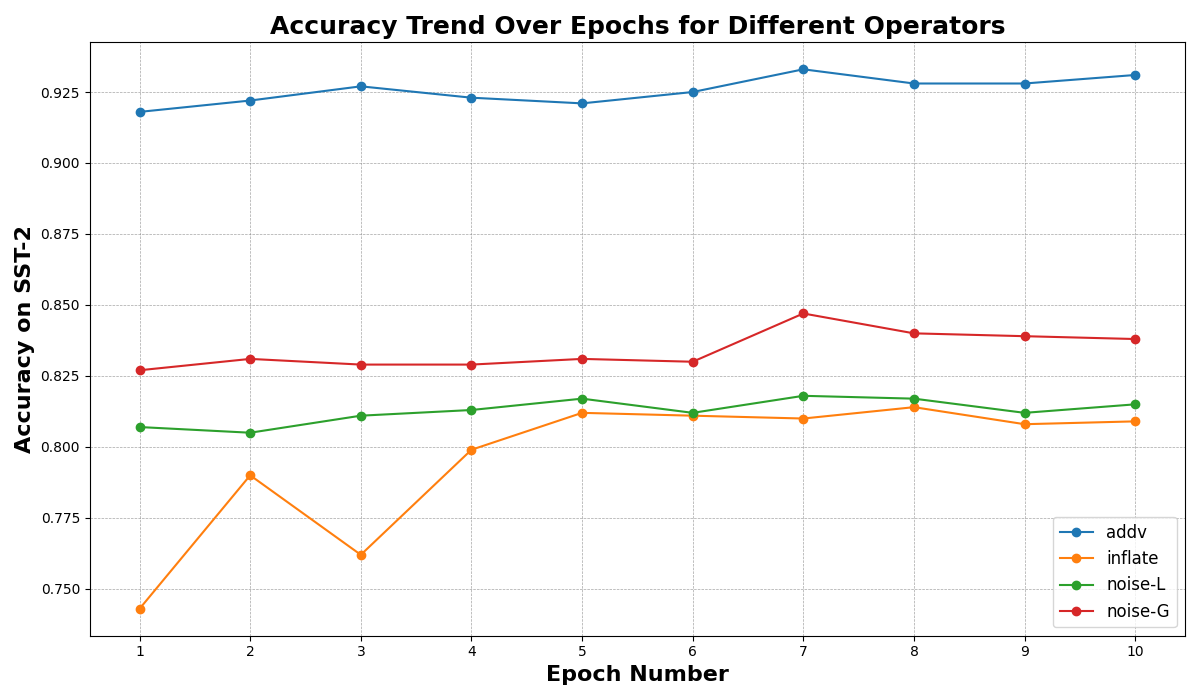}
    \caption{Impact on SST-2 task}
    \label{fig:opts-qqp}
\end{subfigure}
\hfill % this will insert a non-breaking space that will push the subfigures to the edge of the text area
\begin{subfigure}[b]{0.45\textwidth}
    \includegraphics[width=\textwidth]{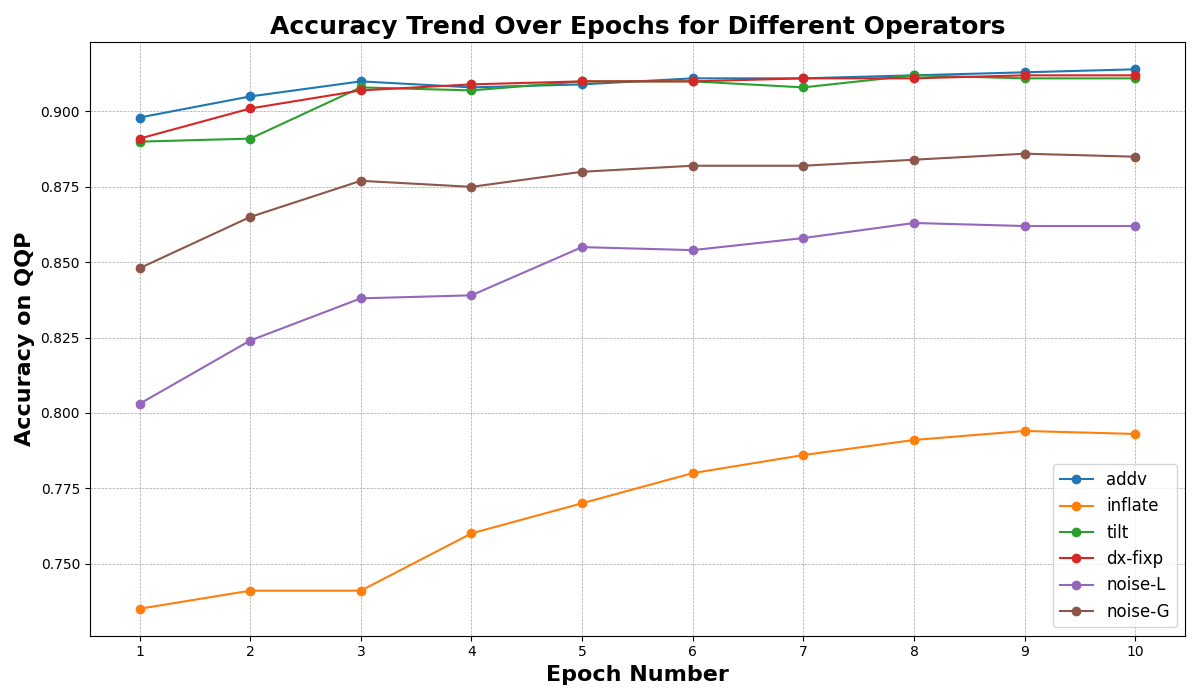}
    \caption{Impact on QQP task}
    \label{fig:opts}
\end{subfigure}
\hfill

\caption{Impact of different noise operators}
\label{fig:impact}

\end{figure}
% From table \ref{tab:implant_results1} and \ref{tab:implant_results2} we can see that with the recovering epoch number increases on the BERT model, the rate of downstream accuracy return grows very little. So for the cost-effectiveness reason, users can choose one epoch as their recovery setting parameter. 

We have performed our experiments with different noise types and multipliers. These noise types are categorized into "Addv", "Inflate", "tilt", "Dx-fixp", "noise-G" and "noise-L". As shown in Figure \ref{fig:opts}, the embedding noise can impact the accuracy of the test utility. Within each noise multiplier, we have chosen a constant value to show the influence of our $vs\_key$ operators. 

When the training epoch increases, the adaption accuracy will increase range from 1.7\% to 4\%. Different operators can impact the accuracy largely with the position and noise type changes inventively. For example, addv type can recover very efficiently in one epoch. Inflate type needs more recover epochs to achieve a reliable accuracy level on SST-2 task.

\subsection{Security Analysis}

\zhou{\textbf{Embedding Inversion Attacks.} In certain scenarios, adversaries might attempt to invert embeddings to extract private information from training models. To explore this, we executed an embedding inversion attack experiment, according to the same setting in \cite{du2023dp}. The results, presented in Table \ref{exp:era}, indicate that higher success rates correspond to a higher risk of attack. Notably, under these embedding inversion attack conditions, CypherTalk demonstrated superior performance, effectively mitigating the risk and showcasing its robustness against such adversarial tactics.  }

 \begin{table}[htbp]
\centering
\small
\caption{Success rates of embedding reverse attack on SST-2, IMDB, and QQP}\label{exp:era}
\begin{tabular}{@{}lccc@{}}
\toprule
& \textbf{SST-2} & \textbf{IMDB} & \textbf{QQP} \\ \midrule
Non-private  & 1         & 1        & 1      \\
DP-SGD       & .9991         & .9982        & 1.      \\
DP-Forward   & .1622         & .1241        & .2226       \\
%MPCFormer    & -              & -             & -            \\
CypherTalk         & .0610         & .0386        & .0433       \\ \bottomrule
\end{tabular}
\end{table}

\begin{table}[htbp]
\centering
\setlength{\tabcolsep}{4pt}
\small
\caption{Sensitive Attribute attack on IMDB}\label{exp:saa}
\begin{tabular}{@{}lccccc@{}}
\toprule
& \textbf{action} & \textbf{comedy} & \textbf{drama} &\textbf{horror} &\textbf{overall}\\ \midrule
DP-SGD& 0.664   & 0.733    & 0.253   & 0.324   &0.536      \\
DP-Forward& 1   & 0    & 0   & 0   &0.276        \\
%MPCFormer& ---  & ---    & ---     & ---    & ---  \\
CypherTalk & 1   & 0    & 0    &0  & 0.276  \\ 
\bottomrule
\end{tabular}
\end{table}

\zhou{\noindent\textbf{Sensitive Attribute Inference Attacks.}  We further investigated sensitive attribute inference attacks, comparing DP-based methods and CypherTalk under identical experimental settings in Du et al. (2023) \cite{du2023dp}. The findings, as illustrated in Table \ref{exp:saa}, reveal that higher values denote increased risk levels. Our analysis indicates that CypherTalk achieves performance comparable to that of DP-Forward, while significantly outperforming DP-SGD. 
}

%Different from cryptography-based methods which have a strong formal verification of mathematics proof, DP-based methods are usually probabilistic in applications which may harm privacy to some extent. Therefore, we analyze sensitive attribute inference attacks between DP-based methods and ours. From the results of \cite{du2023dp,li2021large} we can observe that directly perturb input embeddings can achieve better attribute privacy performance than DP-SGD due to information loss in original data. Compared with the two mentioned methods, ours is more secure since the $HS\_tab$ is only known to the client. If the $HS\_tab$ leaks, the client can require an updated one to replace the previous one. Therefore, the sensitive attributes can be protected in an isolated environment.  

\section{Conclusion}
\zhou{
In this paper, we introduce CypherTalk, a framework that employs cost-effective and self-adaptive shaking and recovery mechanisms. This novel approach is designed to offer robust privacy protection while maintaining high levels of prediction performance. Our evaluations demonstrate the superiority of CypherTalk in comparison to state-of-the-art baselines, highlighting its effectiveness in balancing privacy concerns with accuracy. 

This framework represents an important advancement in the domain, providing a practical solution for privacy-sensitive applications. It enables fine-tuning and deployment of customized LLMs on cloud platforms, all while maintaining optimal performance and data privacy requirements. This approach effectively addresses the critical need for balancing privacy concerns with operational efficacy in the deployment of LLMs.}

\newpage
\section{Limitations}
\zhou{Our approach, while innovative, is not without its limitations. A notable challenge arises when the noise multiplier setting is significantly increased; under such conditions, convergence becomes increasingly difficult to achieve, even when employing a basic affine function as the horizontal shaking function. Furthermore, although CypherTalk demonstrates commendable performance in security analysis experiments, particularly in comparison with crypto-based and DP-based baselines, it lacks a theoretical bound for defense against privacy threats. This absence of a quantifiable protective measure represents a critical aspect where CypherTalk's capability to counter privacy risks needs further exploration and development.}

\bibliography{reference}

\begin{thebibliography}{13}
\expandafter\ifx\csname natexlab\endcsname\relax\def\natexlab#1{#1}\fi

\bibitem[{Boemer et~al.(2019)Boemer, Lao, Cammarota, and Wierzynski}]{boemer2019ngraph}
Fabian Boemer, Yixing Lao, Rosario Cammarota, and Casimir Wierzynski. 2019.
\newblock ngraph-he: a graph compiler for deep learning on homomorphically encrypted data.
\newblock In \emph{Proceedings of the 16th ACM international conference on computing frontiers}, pages 3--13.

\bibitem[{Chen et~al.(2022)Chen, Bao, Huang, Dong, Jiao, Jiang, Zhou, Li, and Wei}]{chen2022x}
Tianyu Chen, Hangbo Bao, Shaohan Huang, Li~Dong, Binxing Jiao, Daxin Jiang, Haoyi Zhou, Jianxin Li, and Furu Wei. 2022.
\newblock The-x: Privacy-preserving transformer inference with homomorphic encryption.
\newblock \emph{arXiv preprint arXiv:2206.00216}.

\bibitem[{Dong et~al.(2023)Dong, Lu, Zheng, Wu, Zhao, Tan, Huang, Hong, Wei, and Cheng}]{dong2023puma}
Ye~Dong, Wen-jie Lu, Yancheng Zheng, Haoqi Wu, Derun Zhao, Jin Tan, Zhicong Huang, Cheng Hong, Tao Wei, and Wenguang Cheng. 2023.
\newblock Puma: Secure inference of llama-7b in five minutes.
\newblock \emph{arXiv preprint arXiv:2307.12533}.

\bibitem[{Du et~al.(2023)Du, Yue, Chow, Wang, Huang, and Sun}]{du2023dp}
Minxin Du, Xiang Yue, Sherman~SM Chow, Tianhao Wang, Chenyu Huang, and Huan Sun. 2023.
\newblock Dp-forward: Fine-tuning and inference on language models with differential privacy in forward pass.
\newblock In \emph{Proceedings of the 2023 ACM SIGSAC Conference on Computer and Communications Security}, pages 2665--2679.

\bibitem[{Gentry(2009)}]{gentry2009fully}
Craig Gentry. 2009.
\newblock Fully homomorphic encryption using ideal lattices.
\newblock In \emph{Proceedings of the forty-first annual ACM symposium on Theory of computing}, pages 169--178.

\bibitem[{Li et~al.(2022)Li, Shao, Wang, Guo, Xing, and Zhang}]{li2022mpcformer}
Dacheng Li, Rulin Shao, Hongyi Wang, Han Guo, Eric~P Xing, and Hao Zhang. 2022.
\newblock Mpcformer: fast, performant and private transformer inference with mpc.
\newblock \emph{arXiv preprint arXiv:2211.01452}.

\bibitem[{Li et~al.(2021)Li, Tramer, Liang, and Hashimoto}]{li2021large}
Xuechen Li, Florian Tramer, Percy Liang, and Tatsunori Hashimoto. 2021.
\newblock Large language models can be strong differentially private learners.
\newblock \emph{arXiv preprint arXiv:2110.05679}.

\bibitem[{Mai et~al.(2023)Mai, Yan, Huang, Yang, and Pang}]{mai2023split}
Peihua Mai, Ran Yan, Zhe Huang, Youjia Yang, and Yan Pang. 2023.
\newblock Split-and-denoise: Protect large language model inference with local differential privacy.
\newblock \emph{arXiv preprint arXiv:2310.09130}.

\bibitem[{Majmudar et~al.(2022)Majmudar, Dupuy, Peris, Smaili, Gupta, and Zemel}]{majmudar2022differentially}
Jimit Majmudar, Christophe Dupuy, Charith Peris, Sami Smaili, Rahul Gupta, and Richard Zemel. 2022.
\newblock Differentially private decoding in large language models.
\newblock \emph{arXiv preprint arXiv:2205.13621}.

\bibitem[{Shamir(1979)}]{shamir1979share}
Adi Shamir. 1979.
\newblock How to share a secret.
\newblock \emph{Communications of the ACM}, 22(11):612--613.

\bibitem[{Wang et~al.(2018)Wang, Singh, Michael, Hill, Levy, and Bowman}]{wang2018glue}
Alex Wang, Amanpreet Singh, Julian Michael, Felix Hill, Omer Levy, and Samuel~R Bowman. 2018.
\newblock Glue: A multi-task benchmark and analysis platform for natural language understanding.
\newblock \emph{arXiv preprint arXiv:1804.07461}.

\bibitem[{Yao(1986)}]{yao1986generate}
Andrew Chi-Chih Yao. 1986.
\newblock How to generate and exchange secrets.
\newblock In \emph{27th annual symposium on foundations of computer science (Sfcs 1986)}, pages 162--167. IEEE.

\bibitem[{Yu et~al.(2021)Yu, Naik, Backurs, Gopi, Inan, Kamath, Kulkarni, Lee, Manoel, Wutschitz et~al.}]{yu2021differentially}
Da~Yu, Saurabh Naik, Arturs Backurs, Sivakanth Gopi, Huseyin~A Inan, Gautam Kamath, Janardhan Kulkarni, Yin~Tat Lee, Andre Manoel, Lukas Wutschitz, et~al. 2021.
\newblock Differentially private fine-tuning of language models.
\newblock \emph{arXiv preprint arXiv:2110.06500}.

\end{thebibliography}
\bibliographystyle{acl_natbib}
% \appendix

% \section{Privacy-aware tuning}
% \label{sec:tuning}
% \begin{figure*}[htb]
%     \centering    
%     \begin{center}
%         \includegraphics[width=1\textwidth]{pics/CypherTalk2.pdf}
%         \caption{Privacy-aware Tuning Process}
%         \label{fig:private fine tuning} 
%     \end{center}
% \end{figure*}
% \begin{figure*}[htb]
%     \centering    
%     \begin{center}
%         \includegraphics[width=1\textwidth]{pics/CypherTalk43.pdf}
%         \caption{Private Inference Process}
%         \label{fig:private inference} 
%     \end{center}
% \end{figure*}
% \section{Embedding Inversion Attack}

% We have performed embedding inversion attack on SST-2, IMDB and QQP tasks.

% As shown in table \ref{tab:embedding}, we can conclude that the embedding inversion attack on our mechanism is hard to achieve when adversaries have no knowledge about the $hs\_key$.

% \section{Attribute Inference Attack}

% We have also performed attribute inference attack on IMDB dataset.

% \label{sec:appendix}

% This is a section in the appendix.

\end{document}